\documentclass[11pt,twoside]{article}
\usepackage{asp2006}
\usepackage{graphics}
\usepackage{epsf}
\usepackage{lscape}
\def\edcomment#1{\iffalse\marginpar{\raggedright\sl#1\/}\else\relax\fi}
\reversemarginpar

\begin{document}
\title{Hanle Effect Diagnostics of the Coronal Magnetic Field\\
{\small{A Test Using Realistic Magnetic Field Configurations}}}
\author{N.-E. Raouafi, S.K. Solanki and T. Wiegelmann}
\affil{National Solar Observatory, 950 N. Cheery Avenue, Tucson, AZ 85719, USA}

\begin{abstract}
Our understanding of coronal phenomena, such as coronal plasma thermodynamics, faces a major
handicap caused by missing coronal magnetic field measurements. Several lines in the UV
wavelength range present suitable sensitivity to determine the coronal magnetic field via the Hanle
effect. The latter is a largely unexplored diagnostic of coronal magnetic fields with a very high
potential. Here we study the magnitude of the Hanle-effect signal to be expected outside the solar
limb due to the Hanle effect in polarized radiation from the H~{\sc{i}} Ly$\alpha$ and $\beta$
lines, which are among the brightest lines in the off-limb coronal FUV spectrum. For this purpose
we use a magnetic field structure obtained by extrapolating the magnetic field starting from
photospheric magnetograms. The diagnostic potential of these lines for determining the coronal
magnetic field, as well as their limitations are studied. We show that these lines, in particular
H~{\sc{i}} Ly$\beta$, are useful for such measurements.
\end{abstract}


\section{Introduction}

The magnetic field is the main driver of all physical phenomena in the solar corona. The field
lines of force thread the solar atmosphere coupling its different layers to the solar interior.
They transport energy to the corona, where it is dissipated and heats the plasma to several MK as
well as accelerating particles to several hundreds of km~s$^{-1}$ within a few solar radii above
the surface. The magnetic field also affects the interplanetary medium through the transfer of
angular momentum from the Sun through the solar wind and transient events. The latter also have
important consequences on planets such as the Earth. In spite of its key role, the magnetic field
remains an almost unknown parameter for any study involving the solar upper atmosphere.

Measuring the magnetic field vector is a routine exercise in the photosphere and to a smaller
degree in the chromosphere. Magnetic fields in these two layers are strong enough to yield an
easily measurable Zeeman signal (and also Hanle effect for numerous spectral lines). Inversion
techniques of the observed Stokes parameters are sufficiently sophisticated from both, a physical
and computational point of view. However, the coronal magnetic field is rather weak, so that the
Zeeman splitting in (generally weak) coronal lines is too small to be accurately measured except
above active regions with relatively strong field for few spectral lines with exceptionally large
Land\'e factors (Lin et al. 2004). Significant Zeeman splitting is usually accompanied by the
so-called ``strong regime of the Hanle effect'' where the direction of the magnetic field projected
on the plane of the sky can be obtained from the linear polarization signal of spectral lines such
as the Fe~{\sc{xiv}} 530.3~nm green line. However, the coronal plasma is optically thin and direct
coronal magnetic field measurements have a line-of-sight integrated character. To obtain the 3D
structure of the magnetic field, vector tomographic methods are required (see Kramar et al. 2006
and Kramar \& Inhester 2007).

Techniques based on different physical mechanisms, such as radio gyroresonant and bremsstrahlung
emission, are being developed. However, they provide field strengths in different layers (depending
on the frequency-opacity relation) without height information on where the signal comes from (see
Lee et al. 1997). They also remain restricted to areas above active regions. These measurements
are, however, important to constrain coronal magnetic fields obtained though the extrapolation of
photospheric and chromospheric field measurements. Extrapolation techniques of magnetic fields have
undergone a rapid development and could benefit enormously from high quality and spatially extended
measurements of magnetic fields in the photosphere and chromosphere and probably the lower corona
(Solanki et al. 2003).

Theoretically, direct determination of the magnetic field in the solar corona could be achieved
through linear polarization of spectral lines with suitable sensitivity to the Hanle effect.
Raouafi et al. (1999a,b) successfully measured the linear polarization of the O~{\sc{vi}} 103.2~nm
line and interpreted it in terms of the Hanle effect signature of the coronal magnetic field (see
also Raouafi et al. 2002a,b).

The aim of the present paper is to study the possibility of measuring the coronal magnetic field
vector through the Hanle effect using coronal lines with both, relatively adequate sensitivity to
the field and high emissivity. The study aims to be relatively realistic in the sense that the
magnetic field used is an extrapolation of photospheric measurements. The results presented here
are in a preliminary stage and the different inputs to the method have to be improved in the future
in order to be used to underpin future missions to measure the coronal magnetic field directly.

\section{Hanle Effect}

The Hanle effect is the result of interferences between different sub-levels of a given atomic
transition. This occurs in resonant scattering conditions when the magnetic field and the upper (or
lower) level of the transition fulfill the following relation
\begin{equation}
\frac{g_{_J}\;\mu_B}{\hbar} \; \tau \; B = 1 ,
\end{equation}
where $B$ is the magnetic field strength, $\tau$ is the lifetime of the atomic level, $g_{_J}$
its Land\'e factor, $\mu_B$ is the Bohr magneton and $\hbar$ the reduced Planck's constant.

\begin{table}[!ht]
\label{fonts}
\caption{Sensitivity to the Hanle effect of H~{\sc{i}} and O~{\sc{vi}} 103.2 nm lines}
\smallskip
\begin{center}
{\small
\begin{tabular}{cccc}
\tableline
\noalign{\smallskip}
Line & $\lambda$ (\AA) & $A_{ul}$ ($10^8$ s$^{-1}$)  & $B_{\rm{Hanle}}$ (Gauss)\\
\noalign{\smallskip}
\tableline
\noalign{\smallskip}
H~{\sc{i}} Ly-$\alpha$  & 1215.16 & 6.265 & 53.43 \\
H~{\sc{i}} Ly-$\beta$   & 1025.72 & 1.672 & 14.26 \\
H~{\sc{i}} Ly-$\gamma$  &  972.53 & 0.682 & 5.81 \\
H~{\sc{i}} Ly-$\delta$  &  949.74 & 0.344 & 2.93 \\
O~{\sc{vi}}            &  1031.91 & 4.160 & 35.48\\
\tableline
\end{tabular}
}
\end{center}
\end{table}

In the solar corona ($B<100$ Gauss), the Hanle effect works better in the ultraviolet and extreme
ultraviolet wavelength ranges (see Table~1). The effect on the linear polarization depends on both
the magnitude and direction of the magnetic field vector. An advantage with respect to the Zeeman
effect is that the spectral resolution is not a critical requirement since the linear polarization
resulting from resonant scattering is of one sign.

\section{Polarization in H~{\sc{i}} Lines}

H~{\sc{i}} Ly$\alpha$ and Ly$\beta$ are both formed by doublets at practically the same wavelength
each (1215.668~{\AA} 1s~$^2{\rm{S}}_{1/2}$~--~2p~$^2{\rm{P}}_{3/2}$ and 1215.674~{\AA}
1s~$^2{\rm{S}}_{1/2}$~--~2p~$^2{\rm{P}}_{1/2}$ for Ly$\alpha$ and 1025.722~{\AA}
1s~$^2{\rm{S}}_{1/2}$~--~3p~$^2{\rm{P}}_{3/2}$ and 1025.723~{\AA}
1s~$^2{\rm{S}}_{1/2}$~--~3p~$^2{\rm{P}}_{1/2}$ for Ly$\beta$). Lifetimes of the components of each
doublet are the same (see Table~1) and the first components can be linearly polarized in resonant
scattering conditions. The second components are not polarizable. These two lines are the most
intense lines emitted in the solar corona, where Ly$\alpha$ is almost exclusively formed by
resonant scattering of the chromospheric radiation (the collisional component contributes less than
1\%). Ly$\beta$ has an important contribution from electron collisions which usually diminishes the
degree of linear polarization of the line. It does not, however, affect the direction of the plane
of polarization.

\section{Model for the Solar Corona}

 \begin{figure}[!h]
 \plottwo{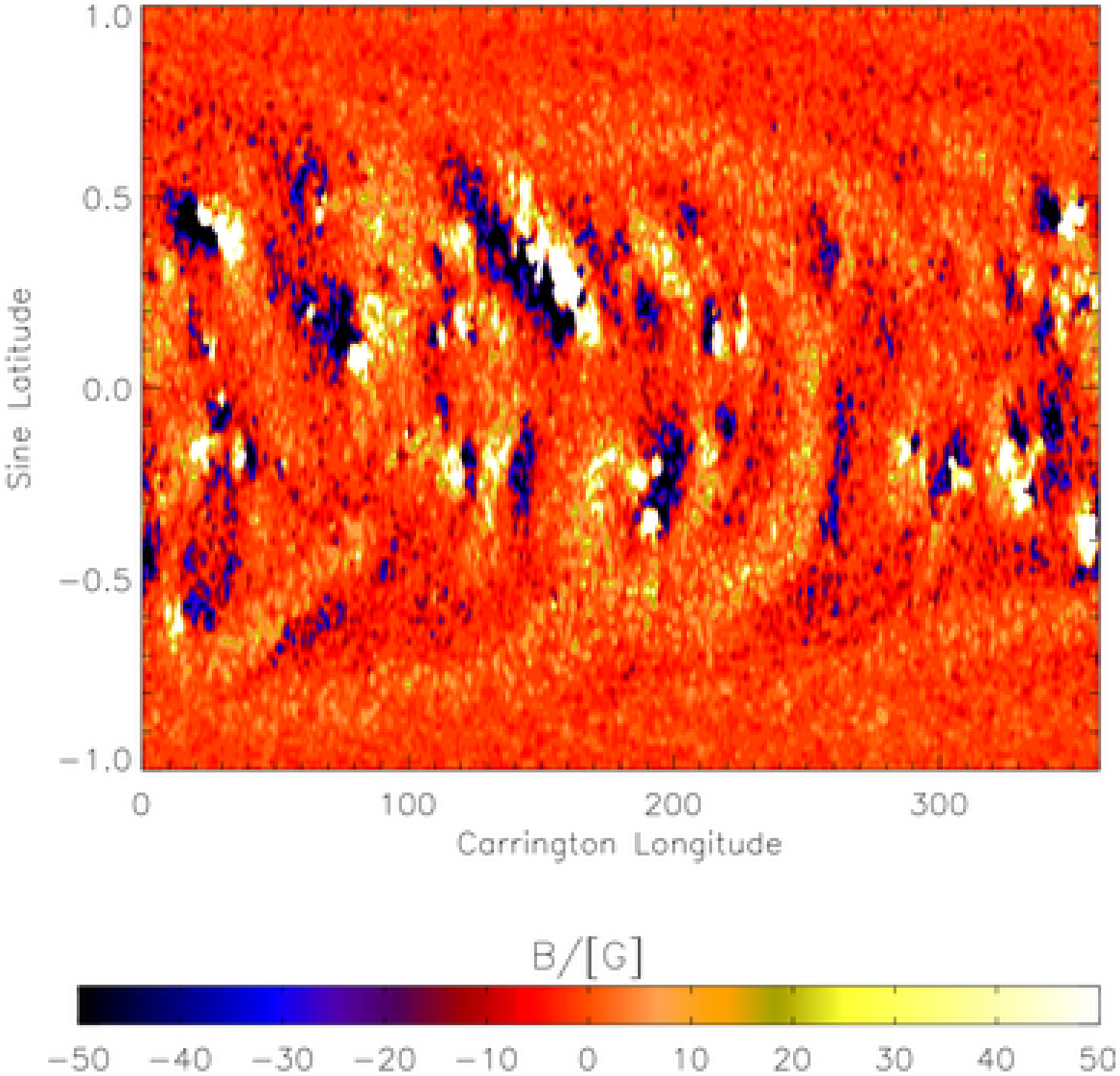}{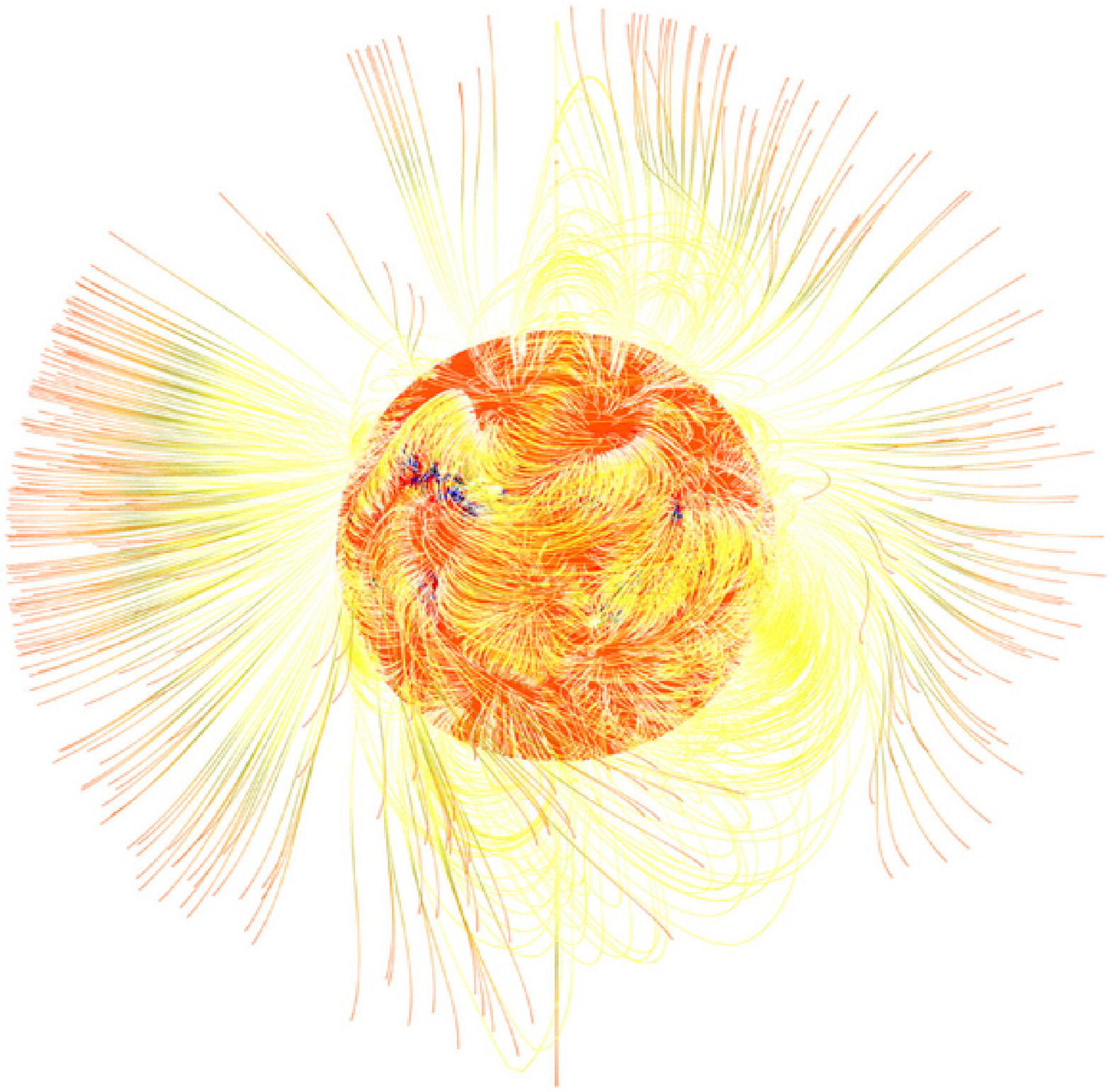}
 \caption{{\itshape Left:\/} Synoptic map of the photospheric magnetic field from SOHO/MDI for
 Carrington rotation 1975.
 {\itshape Right:\/} Coronal field topology obtained from potential field extrapolation.}
 \label{model_corona_mag}
 \end{figure}

As a first step towards a realistic coronal magnetic field model, we consider a coronal magnetic
field distribution obtained from potential field extrapolation of photospheric measurements from
SOHO/MDI. We choose the synoptic map for Carrington rotation number 1975, which corresponds to
maximum activity of solar cycle 23 (see left panel of Figure~\ref{model_corona_mag}). The grid
resolution in spherical coordinates is $40\times80\times160$ in $r$, $\theta$ and $\varphi$
($1~R_{\sun}\le r\le2.5~R_{\sun}$, with a step of $0.0375~R_{\sun}$, $0^\circ\le\theta\le180^\circ$
with a step of $2.5^\circ$, $0^\circ\le\varphi\le360^\circ$ with a step of $2.5^\circ$). Spherical
harmonics up to $l=20$ have been used for computing the coronal magnetic field. The right panel of
the same Figure illustrates the topology of the field lines of force.

 \begin{figure}[!h]
 \plottwo{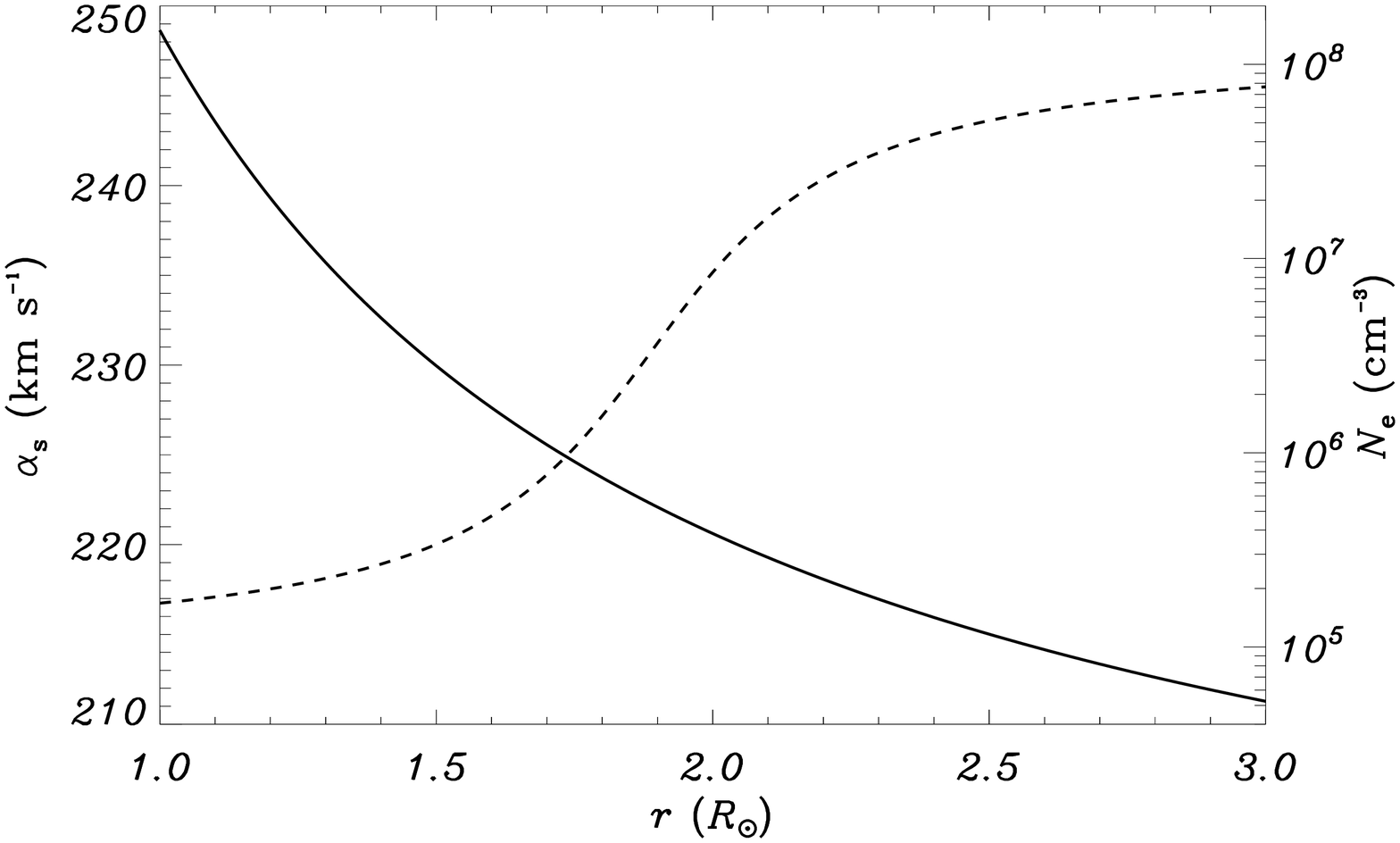}{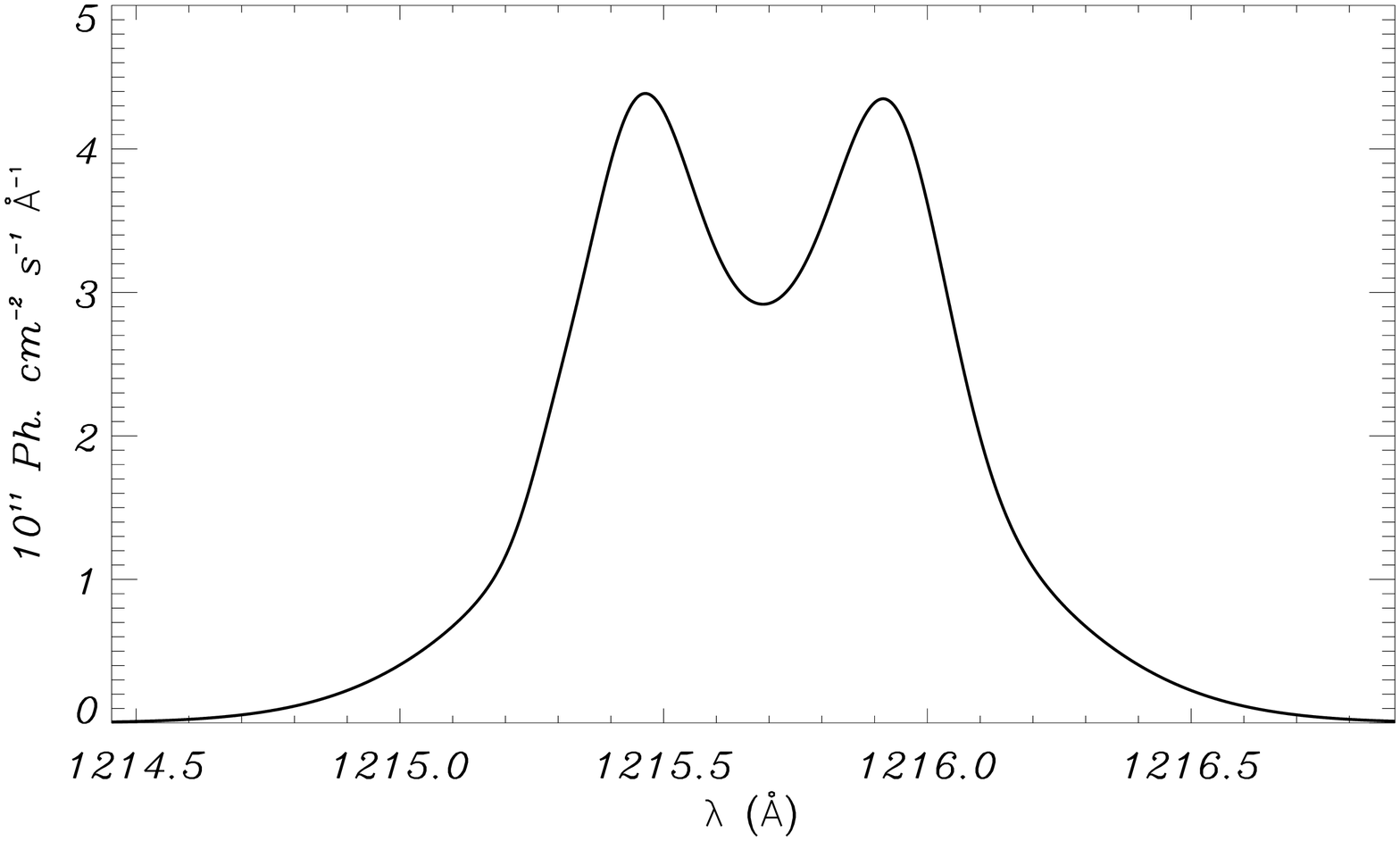}
 \caption{{\itshape Left:\/} Coronal electron density (solid) and velocity turbulence of the
 hydrogen atoms (dashes) as a function of the distance to Sun center.
 {\itshape Right:\/} Chromospheric Ly$\alpha$ profile obtained from SOHO/SUMER (see Lemaire et al.
 1998).}
 \label{model_corona_electur}
 \end{figure}

The electron density model used is from SOHO measurements (left panel of
Figure~\ref{model_corona_electur}; see Doyle et al. 1999). We assume that coronal electron density
depends only on the distance to Sun center. This assumption might be appropriate for coronal
background densities but not for structures such as loops. To address this problem, a detailed
independent study has to be done, which goes beyond the present exploratory study. The coronal
plasma is assumed to be in a static state. This holds adequately at low coronal altitudes in
particular outside of polar coronal holes due to the low speed of the solar wind and also the large
widths of the H~{\sc{i}} chromospheric lines. The temperature of the plasma is assumed to be
isotropic and the velocity  turbulence is given by the left panel of
Figure~\ref{model_corona_electur} (see Raouafi et al. 2007). Integration along the line of sight is
taken into account. We also assume that coronal atoms are excited exclusively by radiation from the
solar disk (see right panel of Figure~\ref{model_corona_electur}; Lemaire et al. 1998). This is
well fulfilled by the case of Ly$\alpha$ but not of Ly$\beta$ which has an important collisional
component. However, we are primarily interested in the direction of the plane of polarization that
is defined only by Stokes parameters $Q$ and $U$, which are created only by radiative excitations.

\section{Results and Discussion}

\begin{figure}[!h]
 \plotone{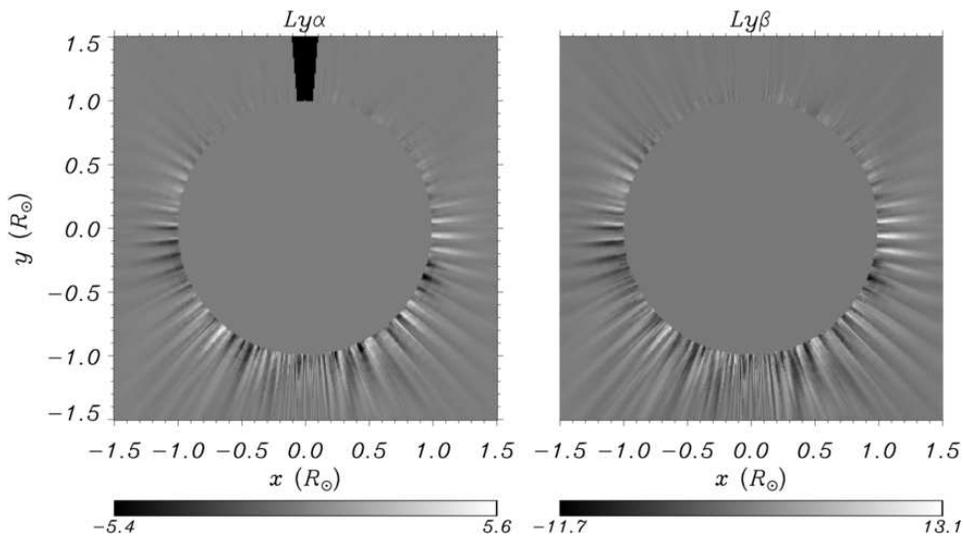}
 \caption{Plane of polarization direction with respect to the tangent to the solar limb for
 H~{\sc{i}} Ly$\alpha$ (left) and $\beta$ (right) obtained from Hanle effect due the magnetic field
 data and the coronal parameters presented in the previous section. Integration along the line of
 sight is taken into account. The grey scales at the bottom indicate the angle of rotation of the
 polarization direction rotation in degrees. The regular (periodic) azimuthal structure is due to
 the limited orders of spherical harmonics (20 first orders) used in the extrapolation of the
 photospheric field. However, significant values of the rotation angle of the plane of polarization
 coincide with flux concentrations as shown in Figure~\ref{model_corona_mag}, which reflect the
 Hanle effect due the magnetic field.} \label{plane_polaraziation}
 \end{figure}

Figure~\ref{plane_polaraziation} displays the direction of the plane of polarization with respect to
the tangent to the solar limb for H~{\sc{i}} Ly$\alpha$ (left) and Ly$\beta$ (right). The angle of
rotation of the plane of polarization is represented by the grey scales plotted below each panel.

Although these results are still preliminary, they suggest that the hydrogen Ly$\alpha$ and $\beta$
lines could be very useful for measuring the coronal magnetic field. Theoretical polarization rates
(not shown here) in these coronal lines are reasonably high and then could be easily gauged.
However, despite the fact that the extrapolation has a significant periodic effect on the outcome
of the calculations, it is also clear that large values of the rotation angle of the polarization
direction have often spatial coherence with concentrations of magnetic field lines as shown by the
top-right panel of Figure~\ref{model_corona_mag}. Generally, the obtained values are well within
the measurement capability of actual polarimeters (say $1^\circ$). This is encouraging indeed,
although numerous aspects in the model could be improved significantly. The present study reveals
also the limits of field extrapolation techniques when a limited number of orders in the solution
are considered. The latter might be sufficient to approximate the field topology to the Extreme
UltraViolet (EUV) coronal structure, but not for quantitative studies such as the present one.

The main problem facing off-limb observations is that such measurements integrate along the line of
sight over structures with different magnetic field strengths and geometries. Even though the
region exactly above the limb (at quadrature) will have the greatest weight, this integration makes
the interpretation difficult.

\section{Conclusion}

Although the model used to compute the polarization of the hydrogen lines is very simple and can be
significantly improved in numerous aspects, significant and practically measurable polarization
parameters (rates and direction angles of the plane of polarization) due to the Hanle effect
resulting from the coronal magnetic field are obtained. The use of coronal magnetic field data
obtained from the potential field extrapolation of photospheric magnetograms is a step forward in
realistically modeling solar coronal magnetic phenomena. Future missions will benefit from
such simulations of the measurement of the coronal magnetic field.

Hanle effect measurements in the UV are a promising way of measuring the Sun's coronal magnetic
field, although due to the lack of appropriate instruments the use of the technique has so far been
limited (e.g. Raouafi et al. 1999a,b). Ly$\alpha$ and $\beta$ are promising spectral lines for
coronal Hanle effect measurements. Their Hanle sensitivity to different field strengths and their
different formation processes make them complementary to each other. The polarization of several
other spectral lines (e.g., O~{\sc{vi}} 1032~{\AA}) have to be explored in order to widen the
choice range for any attempt to measure the magnetic field in the solar corona.

\acknowledgments
The National Solar Observatory (NSO) is operated by the Association of Universities for Research in
Astronomy, Inc., under cooperative agreement with the National Science Foundation. NER's work is
supported by NSO and NASA grant NNH05AA12I. The photospheric data used for the potential field
model are from SOHO/MDI.


\begin{thebibliography}{}

\bibitem[]{} Doyle, J. G., Keenan, F. P., Ryans, R. S. I., Aggarwal, K. M., \& Fludra, A. 1999,
\solphys, 188, 73

\bibitem[Kramar, Inhester, and Solanki(2006)]{2006A&A...456..665K} Kramar,  M., Inhester, B., \&
Solanki, S.K. 2006, \aap, 456, 665

\bibitem[Kramar and Inhester(2007)]{2007MmSAI..78..120K} Kramar, M., \& Inhester, B. 2007, Memorie
della Societa Astronomica Italiana, 78, 120

\bibitem[Lee et al. (1997)]{lee97} Lee, J., White, S. M., Gopalswamy, N., \& Kundu, M. R. 1997,
\solphys, 174, 175

\bibitem[]{} Lemaire, P., Emerich, C., Curdt, W., Schuehle, U., \& Wilhelm, K. 1998, \aap, 334, 1095

\bibitem[Lin et al. (2004)]{lin04} Lin, H., Kuhn, J. R., \& Coulter, R. 2004, \apjl, 613, 177L

\bibitem[]{} Raouafi, N.-E., Lemaire, P., \& Sahal-Br\'echot, S. 1999a, \aap, 345, 999

\bibitem[]{} Raouafi, N.-E., Sahal-Br\'echot, S., Lemaire, P., \& Bommier, V. 1999b, Proc. of the
2$^{\rm{nd}}$ Solar Polarization Workshop. Bangalore-India 12-16 October 1998. Eds. K.N. Nagendra
\& J.O. Stenflo. Boston, Mass.: Kluwer Academic Publishers, 1999. Astrophysics and Space Science
Library, vol. 243, p.349-362

\bibitem[]{} Raouafi, N.-E., Sahal-Br\'echot, S., Lemaire, P., \& Bommier, V. 2002a, \aap, 390, 691

\bibitem[]{} Raouafi, N.-E., Sahal-Br\'echot, S., \& Lemaire, P. 2002b, \aap, 396, 1019

\bibitem[]{} Raouafi, N.-E., Harvey, J. W., \& Solanki, S. K. 2007, \apj, 658, 643

\bibitem[]{} Solanki, S. K., Lagg, A., Woch, J., Krupp, N., \& Collados, M. 2003, \nat, 425, 692

\end{thebibliography}
\end{document}